\documentclass[12pt]{iopart}

\usepackage{iopams}
\usepackage{eucal}
\usepackage{graphicx}

\newcommand{\newc}{\newcommand}
\newc{\beq}{\begin{equation}}
\newc{\eeq}{\end{equation}}
\newc{\kt}{\rangle}
\newc{\bra}{\langle}
\newc{\beqa}{\begin{eqnarray}}
\newc{\eeqa}{\end{eqnarray}}
\newc{\pr}{\prime}
\newc{\longra}{\longrightarrow}
\newc{\ot}{\otimes}
\newc{\rarrow}{\rightarrow}
\newc{\h}{\hat}
\newc{\bom}{\boldmath}
\newc{\btd}{\bigtriangledown}
\newc{\al}{\alpha}
\newc{\be}{\beta}
\newc{\ld}{\lambda}
\newc{\ldmin}{\lambda_{\rm min}}
\newc{\sg}{\sigma}
\newc{\p}{\psi}
\newc{\eps}{\epsilon}
\newc{\om}{\omega}
\newc{\mb}{\mbox}
\newc{\tm}{\times}
\newc{\hu}{\hat{u}}
\newc{\hv}{\hat{v}}
\newc{\hk}{\hat{K}}
\newc{\ra}{\rightarrow}
\newc{\non}{\nonumber}
\newc{\ul}{\underline}
\newc{\hs}{\hspace}
\newc{\longla}{\longleftarrow}
\newc{\ts}{\textstyle}
\newc{\f}{\frac}
\newc{\df}{\dfrac}
\newc{\ovl}{\overline}
\newc{\bc}{\begin{center}}
\newc{\ec}{\end{center}}
\newc{\dg}{\dagger}
\newc{\prh}{\mbox{PR}_H}
\newc{\prq}{\mbox{PR}_q}

\begin{document}

\title{Entangled random pure states with orthogonal symmetry: exact results.}

\author{Pierpaolo Vivo}
\address{Abdus Salam International Centre for
Theoretical Physics, Strada Costiera 11, 34151 Trieste, Italy}
\ead{pvivo@ictp.it}
\begin{abstract}
We compute analytically the density $\varrho_{N,M}(\lambda)$ of Schmidt eigenvalues, distributed according to a fixed-trace Wishart-Laguerre measure, and the average R\'enyi entropy $\langle\mathcal{S}_q\rangle$ for reduced density matrices of entangled random pure states with orthogonal symmetry $(\beta=1)$. 
The results are valid for arbitrary dimensions $N=2k,M$ of the corresponding Hilbert space partitions, and are in excellent agreement with numerical simulations.
\end{abstract}

\maketitle

\section{Introduction} 

Entanglement is one of the most distinctive features of quantum systems. Recently it has attracted much attention 
in view of possible applications to quantum information and quantum computation problems
\cite{NeilsenBook,PeresBook}. In these domains, one is often interested in creating states with large entanglement, thus raising the question how to 
give a quantitative measure of entanglement. Pure bipartite systems (defined below) constitute a typical example
where well-behaved entanglement quantifiers can be defined, such as the von-Neumann or R\'enyi entropies of either subsystem \cite{PeresBook}, the so
called concurrence for two-qubit systems~\cite{Wootters} or other entanglement monotones \cite{cappellini,gour}. 

Introducing a source of randomness in quantum entanglement problems is the key to address \emph{typical} properties
of such states: in this paper we focus on {\it random pure} quantum states in bipartite systems, where many analytical results
have been obtained in recent times (see e.g. \cite{majreview} for an excellent review).

More precisely, we consider a bipartite partition of a $NM$-dimensional Hilbert space ${\cal 
H}^{(NM)}$ as ${\cal H}^{(NM)}={\cal H}^{(N)}_A \otimes {\cal H}^{(M)}_B$, where we assume
without loss of 
generality that $N\le M$.
For example, $A$ may be taken as a given
subsystem (say a set of spins) and $B$ may represent the environment (e.g., a 
heat bath).
Take a quantum state $|\psi\kt$ of the composite system and let $|i^A\kt$ and $|\alpha^B\kt$ be two
complete basis of ${\cal H}^{(N)}_A $ and ${\cal H}^{(M)}_B$ respectively.  The state $|\psi\kt$ can then be expanded as a
linear combination
\begin{equation}\label{psi111}
|\psi\kt= \sum_{i=1}^N\sum_{\alpha=1}^M x_{i,\alpha}\, 
|i^A\kt\otimes |\alpha^B\kt
\end{equation}
whose coefficients $x_{i,\alpha}$'s form the entries of a rectangular $(N\times M)$ matrix $\mathcal{X}$.

Possible features of $|\psi\kt$ we are considering here are the following:
\begin{itemize}
\item {\bf Entanglement:} we say that $|\psi\kt$ is an entangled state if it \emph{cannot} be expressed as a direct 
product
of two states belonging to the two subsystems $A$ and $B$. In other words,
in order for $|\psi\kt$ to be \emph{fully unentangled}, the coefficients $x_{i,\alpha}$ must have the product form
$x_{i,\alpha}= a_i b_{\alpha}$ for all $i$ and $\alpha$ in a certain basis.
In this case,  the state $|\psi \kt= 
|\theta^A\kt \otimes |\theta^B \kt $ can be written as a direct product of two states 
$|\theta^A \kt=\sum_{i=1}^N a_i |i^A\kt$ and 
$|\theta^B\kt= \sum_{\alpha=1}^M b_\alpha |\alpha^B\kt$
belonging respectively to the two subsystems $A$ and $B$.
\item {\bf Randomness:} suppose that the expansion coefficients
$x_{i,\alpha}$ are random variables drawn from a certain probability distribution. 
In this case, we say that $|\psi\kt$ is a \emph{random} state, and here we focus on the 
simplest and most common case where $x_{i,\alpha}$'s
are independent and identically distributed (real or complex) Gaussian variables.   
\item {\bf Purity:} the density matrix of the composite system is simply given by $\rho=|\psi\kt \bra 
\psi|$ with the constraint ${\rm Tr}[\rho]=1$, or equivalently $\bra \psi|\psi\kt=1$. 
Note that the composite system may instead be in a statistically {\it mixed} state, with a density
matrix of the form
\beq
\rho = \sum_{k} p_k\, |\psi_k\kt\, \bra \psi_k|,
\label{dmmixed}
\eeq
where $|\psi_k\kt$'s are the pure states of the composite system and $0\le p_k\le 1$
are the probabilities that the composite system is in the $k$-th pure state, with
$\sum_k p_k=1$. We will not consider this case here, and we refer to \cite{osipov} and references therein for recent results on mixed states.
\end{itemize}

Let $|\psi\kt$ be an entangled pure state of a bipartite quantum system.
Its density matrix can then be straightforwardly expressed as
\beq
\rho = \sum_{i,\alpha}\sum_{j,\beta} x_{i,\alpha}\, x_{j,\beta}^*\, |i^A\kt\bra j^A|\otimes 
|\alpha^B\kt 
\bra\beta^B|,
\label{dem2}
\eeq
where the Roman indices $i$ and $j$ run from $1$ to $N$ and the Greek indices $\alpha$ and 
$\beta$ run from $1$ to $M$. We normalize the pure state $|\psi\kt$ to 
unity so that ${\rm Tr}[\rho]=1$. 

Tracing out the environmental degrees of freedom (i.e., those of subsystem $B$) leads to the definition of the \emph{reduced} density matrix $\rho_A= {\rm Tr}_B[\rho]$:
\beq
\rho_A = {\rm Tr}_B[\rho]=\sum_{\alpha=1}^M \bra \alpha^B|\rho|\alpha^B\kt.
\label{rdm1}
\eeq
Using the expansion in Eq. (\ref{dem2}) one gets
\beq
\rho_A = \sum_{i,j=1}^N \sum_{\alpha=1}^M x_{i,\alpha}\, x_{j,\alpha}^*\, |i^A\kt\bra
j^A|=\sum_{i,j=1}^N W_{ij} |i^A\kt\bra j^A|
\label{rdm2}
\eeq
where $W_{ij}$'s are the entries of the $N\times N$ matrix $\mathcal{W}=\mathcal{X} \mathcal{X}^{\dagger}$.
In analogous way, one could obtain the reduced density matrix $\rho_B={\rm Tr}_A[\rho]$ of the 
subsystem $B$ in terms of the $M\times M$ matrix $\mathcal{W}^\prime=\mathcal{X}^\dagger \mathcal{X}$ and find that $\mathcal{W}$ and $\mathcal{W}^\prime$ share
the same set of nonzero (positive) real eigenvalues $\{\lambda_1,\lambda_2,\ldots,\lambda_N\}$.
In the diagonal basis, one can express $\rho_A$ as
\beq
\rho_A= \sum_{i=1}^N \lambda_i \, |\ld^A_i\kt\, \bra \ld^A_i|
\label{diagA}
\eeq
where $|\ld^A_i \kt$'s are the normalized eigenvectors of $\mathcal{W}=\mathcal{X}\mathcal{X}^{\dagger}$ and similarly for $\rho_B$.
The original composite state $|\psi\kt$ in this diagonal basis reads:
\beq
|\psi\kt = \sum_{i=1}^{N} \sqrt{\ld_i}\, |\ld_i^A\kt \otimes |\ld^B_i \kt
\label{Sch1}
\eeq
Eq. (\ref{Sch1}) is known as the Schmidt decomposition, and the 
normalization
condition $\bra \psi|\psi\kt=1$, or equivalently ${\rm Tr}[\rho]=1$, imposes
a constraint on the eigenvalues, $\sum_{i=1}^N \ld_i=1$. 

It is useful to remark that while each individual state $|\ld_i^A\kt \otimes|\ld^B_i \kt$ in the Schmidt decomposition
in Eq. (\ref{Sch1}) is unentangled, their linear combination $|\psi\kt$, in general,
is entangled, and therefore the state $|\psi\kt$ cannot, in general, 
be written as a direct product $|\psi\kt= |\phi^A\kt \otimes |\phi^B\kt$ of two states of the 
respective subsystems. Knowledge of the eigenvalues $\{\lambda_1,\lambda_2,\ldots, \lambda_N\}$ of the matrix $\mathcal{W}$
is essential in providing information
about how entangled a pure state is. Typical entanglement quantifiers include the R\'enyi entropy of order $q\geq 1$
\begin{equation}
\mathcal{S}_q:=\frac{1}{1-q}\log\left[\sum_{i=1}^N \lambda_i^q\right]
\end{equation}
which converges to the von Neumann entropy 
$\mathcal{S}_{\mathrm{VN}}=-\sum_{i=1}^N \lambda_i\ln\lambda_i$ for $q\to 1$. The R\'enyi and von Neumann entropies
attain their minimum value $0$ when one of the eigenvalues reaches its maximum value $1$ and all the others are zero, which corresponds to completely unentangled states,
while they attain their maximum value $\ln N$ in the situation where all eigenvalues are equal ($\lambda_i=1/N$ for all $i$). In this case, all the states in the Schmidt decomposition
(\ref{Sch1}) are equally present and the state $|\psi\kt $ is maximally entangled.

So far, we have considered an arbitrary pure state in Eq. (\ref{psi111}) with
fixed coefficient matrix $\mathcal{X}=[x_{i,\alpha}]$. This state is called random if
the coefficients are drawn from an underlying Gaussian distribution (real or complex)
${\rm Prob}[\mathcal{X}]\propto \exp\left[-\frac{\beta}{2} {\rm Tr}(\mathcal{X}^{\dagger} \mathcal{X})\right]$
where
the Dyson index $\beta=1,2$ corresponds respectively to real and complex 
$\mathcal{X}$ matrices. While generally $\{x_{i,\alpha}\}$ are complex, {\em real} coefficients are important for systems enjoying a time-reversal (or any anti-unitary) symmetry.
In these cases, it is known that the eigenfunctions can be chosen to be real, and the corresponding ensembles are the 'orthogonal' ones $(\beta=1)$. Exact results for the statistics of random orthogonal 
states are very scarce \cite{majbohi,chen}. It is the goal of this paper to fill this gap and to present exact results for the average density of Schmidt eigenvalues (one-point function)
and the average R\'enyi entropy, valid for arbitrary dimensions $N=2k,M$ of the corresponding Hilbert space partitions.

Conversely, analytical results for spectral statistics of random pure states with broken time-reversal symmetry $(\beta=2)$ abound. 
The joint probability density (jpd) of Schmidt eigenvalues was derived by Lloyd and Pagels \cite{LP} (see eq. (\ref{jpddelta}) below), and using this result
Page \cite{Page95} computed the average von Neumann entropy for large $N,M$ and found:
\begin{equation}\label{gg1}
\langle \mathcal{S}_{\mathrm{VN}} \rangle \approx \ln (N)-\frac{N}{2M}
\end{equation}
Since $\ln (N)$ is the maximal possible value of von Neumann entropy for the subsystem $A$, in the limit when $M\gg N$, the average entanglement entropy of a random pure 
state is
close to maximal\footnote{Note, however, that the probability of the maximally entangled microscopic state (where all Schmidt eigenvalues are close to each other) decays very quickly
as $N$ increases, a result that is based on the exact evaluation of the full large deviation tails \cite{majnadal}.}. Later, the same result was shown to hold for the $\beta=1$ case~\cite{Arul1}. In the same paper, Page also conjectured from numerical experiments that the average von Neumann entropy
for {\em finite} $N,M$ and $\beta=2$ should read 
\begin{equation}
\langle \mathcal{S}_{\mathrm{VN}}\rangle = \sum_{k=N+1}^{MN}\frac{1}{k}-\frac{M-1}{2N},
\end{equation}
a result that was independently proven by many researchers soon after \cite{proofspage} also in a non-extensive setting \cite{malacarne}. 
Recently, many efforts have been directed towards the study of other statistical quantities for finite $(N,M)$, and {\em full distributions} of interesting observables.
We mention here:
\begin{itemize}
\item Density of Schmidt eigenvalues (one-point function) for $\beta=2$ and finite $(N,M)$, derived independently in \cite{densitybeta2} and \cite{adachi};
\item Universality of eigenvalue correlations for $\beta=2$ \cite{liu};
\item Distribution of minimum eigenvalue for $\beta=1,2$ and finite $(N,M)$, derived in \cite{majbohi} where a conjecture by Znidaric \cite{Znd} was proven (see also \cite{chen} for a related result);
\item Average fidelity between quantum states \cite{zyc} and distribution of so-called $G$-concurrence \cite{cappellini} for $\beta=2$;
\item Distribution of so-called {\em purity} (i.e. $\mathcal{S}_2$) for small $N$ \cite{giraud}, and phase transitions in its Laplace transform for large $N$ \cite{scard1};
\item Full distribution of R\'enyi entropies (including large deviation tails), computed in \cite{majnadal} for large $N=M$ and all $\beta$s using a Coulomb gas method.
As a byproduct, the authors also obtain in \cite{majnadal} the average and variance of R\'enyi entropy valid for large $N=M$ as
\footnote{Note that the limit $q\to 1$ of eq. (\ref{gg}) is consistent with eq. (\ref{gg1}) already derived by Page.}:
\begin{eqnarray}
\langle\mathcal{S}_q\rangle &\approx\ln N-\frac{\ln\bar{s}(q)}{q-1}\stackrel{q\to 1}{\to}\ln N-\frac{1}{2}\label{gg}\\
\mathrm{Var}(\mathcal{S}_q) &\approx \frac{q}{2\beta N^2}
\end{eqnarray}
where:
\begin{equation}
\bar{s}(q)=\frac{4^q \Gamma(q+1/2)}{\sqrt{\pi}\Gamma(q+2)}
\end{equation}
\end{itemize}
We will compare in Section \ref{aver} the asymptotic result (\ref{gg}) with our exact formula for the average $\langle\mathcal{S}_q\rangle$ for $\beta=1$ (see eq. (\ref{exactbeta1}))
and find that (\ref{exactbeta1}) converges to (\ref{gg}) very quickly for low $q$, thus including the most relevant cases $q=1,2$. Conversely, the rate of convergence
progressively deteriorates as $q$ increases (see Section \ref{aver}).

In order to proceed, we now summarize the basic ingredients of the calculation. The joint distribution of Schmidt eigenvalues $\lambda_i\in [0,1]$ \cite{LP} reads:
\begin{equation}\label{jpddelta}
\mathcal{P}(\lambda_1,\ldots,\lambda_N)=C_{N,M}^{(\beta)}\delta\left(\sum_{i=1}^N\lambda_i -1\right)\prod_{i=1}^N\lambda_i^{\frac{\beta}{2}(M-N+1)-1}\prod_{j<k}|\lambda_j-\lambda_k|^\beta
\end{equation}
where $C_{N,M}^{(\beta)}$ is a normalization constant known exactly for any $\beta$, and the Dyson index $\beta=1,2$ identifies respectively systems with preserved (orthogonal) or broken (unitary) 
time-reversal symmetry. The delta function guarantees that ${\rm Tr}[\rho]=1$ and implies that the typical eigenvalue scales as $\lambda\sim 1/N$. 

Another jpd of eigenvalues which is closely related to (\ref{jpddelta}) is from the Wishart-Laguerre ensemble of
random matrices \cite{Wishart,Mehta} of the form $\mathcal{W}=\mathcal{X}^\dagger\mathcal{X}$, where $\mathcal{X}$ is a Gaussian $M\times N$ matrix with real or complex entries.
The joint distribution of the $N$ nonnegative eigenvalues of $\mathcal{W}$ is known~\cite{James}
\beq
\fl\mathcal{P}^{(W)}(\lambda_1,\ldots,\lambda_N) =[K_{N,M}^{(\beta)}]^{-1}\, e^{-\frac{\beta}{2}\sum_{i=1}^N 
\lambda_i}\, \prod_{i=1}^N \lambda_i^{\frac{\beta}{2}(1+M-N)-1}\, \prod_{j<k} 
|\lambda_j-\lambda_k|^{\beta}
\label{wishart1}
\eeq
where $K_{N,M}^{(\beta)}$ is a known normalization constant. Therefore, the jpd (\ref{jpddelta}) can be seen as a {\em fixed-trace} version of the Wishart-Laguerre ensemble. 
The presence of a fixed-trace
constraint has crucial consequences on the spectral properties of random matrix ensembles \cite{akemann,ASOS}. The goal of this paper is to compute 
exactly the one-point marginal (average density) $\varrho_{N,M}(x)$ for $\beta=1$, defined as:
\begin{equation}
\varrho_{N,M}(x)=\Big\langle \frac{1}{N}\sum_{i=1}^N \delta(x-\lambda_i)\Big\rangle
\end{equation}
where the average $\langle\cdot\rangle$ is taken with respect to the measure (\ref{jpddelta}). Writing down this average explicitly, one is led to:
\begin{equation}\label{density}
\fl\varrho_{N,M}(\lambda_1):=C_{N,M}^{(1)}\int_{[0,1]^{N-1}}d\lambda_2\cdots d\lambda_N \delta\left(\sum_{i=1}^N\lambda_i -1\right)\prod_{i=1}^N\lambda_i^{\frac{\nu -1}{2}}\prod_{j<k}|\lambda_j-\lambda_k|
\end{equation}
where $\nu=M-N$.
Computing this $(N-1)$-fold integral is the main technical challenge. Note that in the large $N,M$ limit with $c=N/M$ fixed, the average density
can be computed for all $\beta$s using a Coulomb gas technique \cite{majnadal} and has the scaling form:
\begin{equation}\label{averagec}
\varrho_{c}(x)=N\varrho_c^\star (Nx)
\end{equation}
where:
\begin{equation}
\varrho_c^\star (x):=\frac{1}{2\pi c x}\sqrt{(L_{(+)}(c)-x)(x-L_{(-)}(c))}
\end{equation}
where the edge points $L_{(\pm)}=c(\sqrt{1/c}\pm 1)^2$.

The average density is important in order to obtain averages of so-called linear statistics\footnote{A linear statistics
is a quantity of the form $\mathcal{O}=\sum_{i=1}^N f(\lambda_i)$, where $f(x)$ is any smooth function.} with a simple
one-dimensional integration as:
\begin{equation}\label{linearstat}
\Big\langle\sum_{i=1}^N f(\lambda_i)\Big\rangle = N\int_0^1 d\lambda\ \varrho_{N,M}(\lambda) f(\lambda)
\end{equation}
In particular, we consider $f(x)=x^q$ for the R\'enyi entropy in Section \ref{aver}.

The paper is organized as follows. In Section \ref{densitysec}, we compute the one-point density for $\beta=1$ and any $N=2k,M$ using a Laplace transform method,
which is technically transparent and avoids the unnecessarily heavy formalism used for earlier derivations of the $\beta=2$ case \cite{adachi}.
In section \ref{aver} we use the obtained result to compute the average R\'enyi entropies for $\beta=1$ and compare them 
with the asymptotic formula (\ref{gg}) for large $N$. Then we provide some conclusions in Section \ref{concl}.

\section{Density of Schmidt eigenvalues for the orthogonal case $\beta=1$ ({\bf $N$ even})}\label{densitysec}
In this case the jpd (\ref{jpddelta}) of Schmidt eigenvalues reads:
\begin{equation}
\mathcal{P}(\lambda_1,\ldots,\lambda_N)=C_{N,M}\delta\left(\sum_{i=1}^N\lambda_i -1\right)\prod_{i=1}^N\lambda_i^{\frac{\nu -1}{2}}\prod_{j<k}|\lambda_j-\lambda_k|
\end{equation}
where we put $C_{N,M}\equiv C_{N,M}^{(1)}$ and $\nu=M-N$.

The goal is to compute the density of eigenvalues for finite $(N,M)$, i.e. the marginal 
\begin{equation}\label{density}
\fl\varrho_{N,M}(\lambda_1):=C_{N,M}\int_{[0,1]^{N-1}}d\lambda_2\cdots d\lambda_N \delta\left(\sum_{i=1}^N\lambda_i -1\right)\prod_{i=1}^N\lambda_i^{\frac{\nu -1}{2}}\prod_{j<k}|\lambda_j-\lambda_k|
\end{equation}
which is normalized to $1$, i.e. $\int_0^1 dx \varrho_{N,M}(x)=1$. 
In the orthogonal case the normalization constant reads $C_{N,M}=\frac{\Gamma(NM/2)(\sqrt{\pi}/2)^N}{\prod_{j=0}^{N-1}\Gamma((M-j)/2)\Gamma(1+(N-j)/2)}$.

We first define $\varrho_{N,M}(\lambda_1)=\hat\varrho_{N,M}(\lambda_1,1)$, where
\begin{equation}\label{densitytbeta1}
\fl\hat\varrho_{N,M}(\lambda_1,t):=C_{N,M}\int_{[0,1]^{N-1}}d\lambda_2\cdots d\lambda_N \delta\left(\sum_{i=1}^N\lambda_i -t\right)\prod_{i=1}^N\lambda_i^{\frac{\nu -1}{2}}\prod_{j<k}|\lambda_j-\lambda_k|
\end{equation}
is an auxiliary function that we are going to compute exactly.

We next take the Laplace transform of (\ref{densitytbeta1}):
\begin{equation}\label{Laplace}
\fl\int_0^\infty dt\hat\varrho_{N,M}(\lambda_1,t)e^{-st}=C_{N,M}\int_{[0,\infty)^{N-1}} d\lambda_2\ldots d\lambda_N e^{-s\sum_{i=1}^N\lambda_i}\prod_{i=1}^N\lambda_i^{\frac{\nu -1}{2}}\prod_{j<k}|\lambda_j-\lambda_k|
\end{equation}
where in the r.h.s. we have extended the range of integration to the full positive semiaxis. This is harmless in view of the unit norm constraint. The integral on the r.h.s. can be written in the form:
\begin{equation}\label{casin}
\fl\int_{[0,\infty)^{N-1}} d\lambda_2\ldots d\lambda_N e^{-s\sum_{i=1}^N\lambda_i}\prod_{i=1}^N\lambda_i^{\frac{\nu -1}{2}}\prod_{j<k}|\lambda_j-\lambda_k|=
\frac{K_{N,M}}{N(2s)^{-1+MN/2}}\varrho_{N,M}^{(\rm WL)}(2s\lambda_1)
\end{equation}
where:
\begin{eqnarray}
\nonumber K_{N,M}&\equiv K_{N,M}^{(1)}=\int_{[0,\infty)^{N}} d\lambda_1\ldots d\lambda_N e^{-\frac{1}{2}\sum_{i=1}^N\lambda_i}\prod_{i=1}^N\lambda_i^{\frac{\nu -1}{2}}\prod_{j<k}|\lambda_j-\lambda_k|=\\
&=2^{NM/2}(\sqrt{\pi}/2)^{-N}\prod_{j=0}^{N-1}\Gamma\left(\frac{3+j}{2}\right)\Gamma\left(\frac{M-N+j+1}{2}\right)
\end{eqnarray}
is the normalization constant of the jpd of eigenvalues of a Wishart-Laguerre (WL) ensemble with $\beta=1$ (which can be derived from the Laguerre-Selberg integral (see e.g. \cite{luque}), and
\begin{equation}
\fl\varrho_{N,M}^{(\rm WL)}(\lambda_1)=N(K_{N,M})^{-1}\int_{[0,\infty)^{N-1}} d\lambda_2\ldots d\lambda_N e^{-\frac{1}{2}\sum_{i=1}^N\lambda_i}\prod_{i=1}^N\lambda_i^{\frac{\nu -1}{2}}\prod_{j<k}|\lambda_j-\lambda_k|
\end{equation}
is the one-point density of the WL ensemble, normalized to $N$.

The spectral density (one-point function) for the WL ensemble $\varrho_{N,M}^{(\rm WL)}(\lambda_1)$ at even $N$ is known:
\begin{equation}\label{finiteNdensitybeta1}
\fl\varrho_{N,M}^{(\rm WL)}(x)=\frac{\theta(x)}{4}x^{(\nu-1)/2}e^{-x/2}\int_0^\infty dx^\prime \mathrm{sgn}(x-x^\prime)(x^\prime)^{(\nu-1)/2}e^{-x^\prime/2}S(x,x^\prime;\nu,N)
\end{equation}
where:
\begin{equation}
S(x,x^\prime;\nu,N):=\sum_{j=0}^{N-2}\frac{(j+1)!}{(j+\nu)!}\{L_{j+1}^\nu (x^\prime)L_j^\nu (x)-L_{j+1}^\nu (x)L_j^\nu(x^\prime)\}
\end{equation}
where $L_j^\nu(x)$ are Laguerre functions defined by the sum: 
\begin{equation}\label{sumlaguerre}
L_N^\nu(z)=\frac{\Gamma(\nu+N+1)}{N!}\sum_{k=0}^N \frac{(-N)_k}{k!\ \Gamma(\nu+k+1)}z^k
\end{equation}
(where $(x)_n=\Gamma(x+n)/\Gamma(x)$) and $\mathrm{sgn}(z)=z/|z|$. The explicit formula (\ref{finiteNdensitybeta1}) can be most conveniently derived by taking the $\mu\to 0 $ limit of eq. 4.14 in 
\cite{akephillips}. Equivalent but less handy expressions can be found in \cite{verb}, while the general formalism based on skew-orthogonal polynomials is in \cite{Mehta,nagao}.

In order to take the inverse Laplace transform of (\ref{casin}), some work is needed. First, we make a change of variable $x^\prime = xz$ in (\ref{finiteNdensitybeta1}),
obtaining:
\begin{equation}\label{finiteNdensitybeta1bis}
\fl\varrho_{N,M}^{(\rm WL)}(x)=\frac{\theta(x)}{4}x^{\nu}e^{-x/2}\int_0^\infty dz\ \mathrm{sgn}(1-z) z^{(\nu-1)/2}e^{-xz/2}S(x,xz;\nu,N)
\end{equation}

For later convenience, we now define and compute the following inverse Laplace transform:
\begin{equation}\label{defPsi}
\Psi_k(t,x,z;N_1,N_2;\nu):=\mathcal{L}^{-1}\left[s^k e^{-sx(1+z)}L_{N_1}^{\nu}(2sx)L_{N_2}^{\nu}(2sxz)\right](t)
\end{equation}
Using the general definition of Laguerre functions (\ref{sumlaguerre}) and the following elementary Laplace inverse:
\begin{equation}
\mathcal{L}^{-1}[s^a e^{-bs}](t)=\frac{(t-b)^{-1-a}\theta(t-b)}{\Gamma(-a)}
\end{equation}
(where $\theta(x)$ is the Heaviside step function), it is straightforward to get:
\begin{eqnarray}
\nonumber &\Psi_k(t,x,z; N_1,N_2;\nu) =\frac{\Gamma(\nu+N_1+1)\Gamma(\nu+N_2+1)}{N_1! N_2!}\times\\
\nonumber &\times\sum_{m=0}^{N_1}\sum_{\ell=0}^{N_2}
\frac{(-N_1)_m (-N_2)_\ell }{m!\ell! \Gamma(\nu+m+1)\Gamma(\nu+\ell+1)\Gamma(-k-m-\ell)}\times\\
& (2x)^{m+\ell}z^\ell \left(t-x(1+z)\right)^{-k-m-\ell-1}\theta\left(t-x(1+z)\right)
\end{eqnarray}

Combining everything together, we obtain for $\hat\varrho_{N,M}(\lambda_1,t)$:
\begin{eqnarray}
\nonumber\hat\varrho_{N,M}(\lambda_1,t) &=
\frac{C_{N,M}K_{N,M}}{ 2^{1+MN/2}N}(2\lambda_1)^\nu\sum_{j=0}^{N-2}\frac{(j+1)!}{(j+\nu)!}\int_0^\infty dz\ \mathrm{sgn}(1-z) z^{(\nu-1)/2}\times\\
&\times\left[
\Psi_\kappa(t,\lambda_1,z; j,j+1;\nu) -\Psi_\kappa(t,\lambda_1,z; j+1,j;\nu) \right]
\end{eqnarray}
where:
\begin{equation}
\kappa= M-N-MN/2+1
\end{equation}

The sought density $(t=1)$ can then be written in the compact form:
\begin{eqnarray}
\nonumber &\varrho_{N,M}(x) =\mathcal{N}_{N,M}\sum_{j=0}^{N-2}\sum_{m=0}^j\sum_{\ell=0}^{j+1} \mathbf{c}_{\ell m}^{(j)}x^{\nu+m+\ell}\times \\
&\nonumber\times\int_0^\infty dz\ \mathrm{sgn}
(1-z)\ z^{(\nu-1)/2}\left(1-x(1+z)\right)^{-\kappa-m-\ell-1}\mathbf\phi_{\ell m}(z)\theta(1-x(1+z))
\end{eqnarray}

where:
\begin{eqnarray}
\mathcal{N}_{N,M} &= \frac{C_{N,M}K_{N,M}}{ 2^{1+MN/2-\nu}N}\\
\mathbf{c}_{\ell m}^{(j)} &= \frac{\Gamma(\nu+j+2) 2^{m+\ell} (-j)_m (-j-1)_\ell}{j! m!\ell! \Gamma(\nu+m+1)\Gamma(\nu+\ell+1)\Gamma(-\kappa-m-\ell)}\\
\mathbf\phi_{\ell m}(z) &=  z^\ell-z^m
\end{eqnarray}

The integral in $z$ can now be performed exactly. Let
\begin{equation}
\Xi(a,b;x)=\int_0^{(1-x)/x} dz\ \mathrm{sgn}(1-z) z^a (1-x(1+z))^b,\qquad \mathrm{Re}[a,b]>-1
\end{equation}
One has:
\begin{equation*}
\fl\Xi(a,b;x) = x^{-1-a} (1-x)^{1+a+b}\times
\cases{
\mathrm{B}(a+1,b+1), \quad\mbox{for } 1/2 \leq x \leq 1\\
-\mathrm{B}(a+1,b+1)+2\mathrm{B}(x/(1-x),a+1,b+1),\\
\hspace{92pt}\mbox{for } 0\leq x\leq 1/2}
\end{equation*}
where $\mathrm{B}(a,b)=\Gamma(a)\Gamma(b)/\Gamma(a+b)$ is Euler's Beta function and $\mathrm{B}(z,a,b)=\int_0^z du\ u^{a-1} (1-u)^{b-1}$
is the incomplete Beta function.

Eventually one gets for the density of Schmidt eigenvalues for $\beta=1$:

\begin{eqnarray}
\nonumber \varrho_{N,M}(x) &=\mathcal{N}_{N,M}\sum_{j=0}^{N-2}\sum_{m=0}^j\sum_{\ell=0}^{j+1} \mathbf{c}_{\ell m}^{(j)}x^{\nu+m+\ell}\left[\Xi\left(\frac{\nu-1}{2}+\ell,-\kappa-m-\ell-1;x\right)\right.\\ &\left.-\Xi\left(\frac{\nu-1}{2}+m,-\kappa-m-\ell-1;x\right)\right],\qquad 0\leq x\leq 1 \label{rhofinale}
\end{eqnarray}
Equation (\ref{rhofinale}) is the main result of this section\footnote{Following analogous but much quicker steps, one can also derive the already known one-point density for $\beta=2$
in a much simpler way. For example, for $N=M$ we obtain
\begin{equation*}\label{finalrho}
\varrho_{N,N}^{(\beta=2)}(x)=\frac{\Gamma(N^2)}{N}\sum_{k=0}^{N-1}\sum_{\ell,m=0}^k\frac{(-k)_\ell\ (-k)_m}{(\ell!)^2\ (m!)^2}\frac{x^{\ell+m}(1-x)^{N^2-2-\ell-m}}{\Gamma(N^2-1-\ell-m)},
\qquad 0\leq x\leq 1
\end{equation*}}. 
The obtained exact formula is the starting point to compute averages of linear statistics using formula (\ref{linearstat}).
In the next section, we are going to compute the average R\'enyi entropy at finite $N,M$ and compare it with the exact asymptotic result for large $N=M$
obtained in \cite{majnadal}. Note that $\langle\lambda\rangle=\int_0^1 d\lambda\ \lambda \varrho_{N,M}(\lambda)=1/N$ in agreement with the general scaling argument
that typically $\lambda\sim 1/N$ due to the trace constraint $\sum_{i=1}^N \lambda_i=1$.

\begin{figure}[htb]
\begin{center}
\includegraphics[bb =0 0 240 161, width=0.7\textwidth]{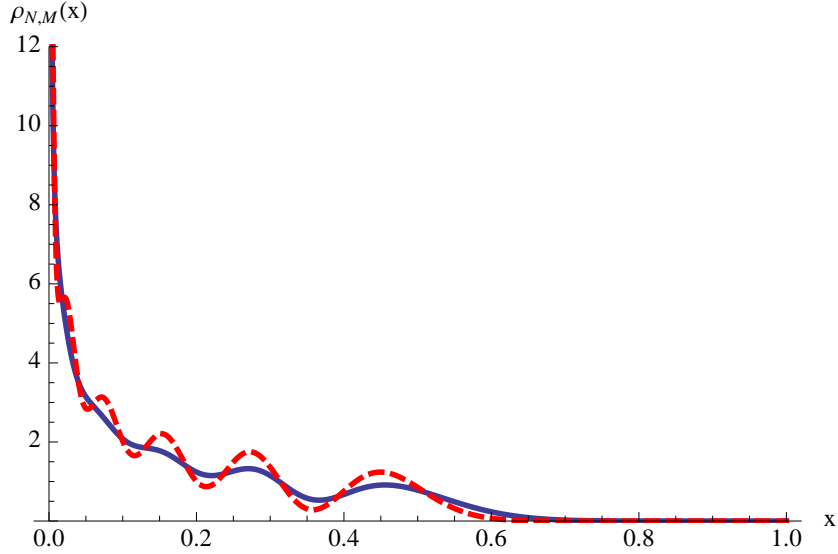}
\caption{Density of Schmidt eigenvalues for $\beta=1$ (solid blue line, eq. (\ref{rhofinale})) and $\beta=2$ (dashed red line), both for $N=M=6$.\label{Bothbetas}}
\end{center}
\end{figure}

\begin{figure}[htb]
\begin{center}
\includegraphics[bb =0 0 240 161, width=0.7\textwidth]{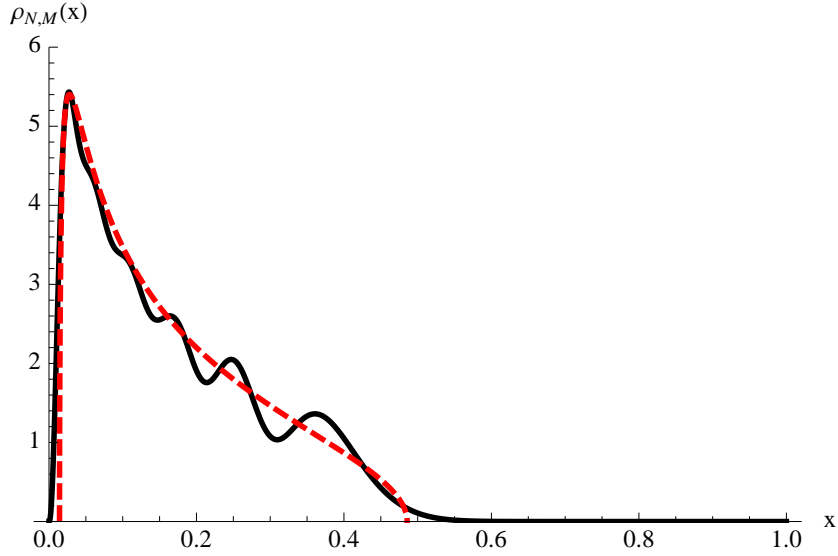}
\caption{Density of Schmidt eigenvalues for $\beta=1$ (eq. (\ref{rhofinale})) for $N=6,M=12$ (solid line). In dashed red, the corresponding large $N$ density (\ref{averagec}) for $c=N/M=1/2$.\label{Beta1}}
\end{center}
\end{figure}

\begin{figure}[htb]
\begin{center}
\includegraphics[bb = -269   112   882   680, width=0.9\textwidth]{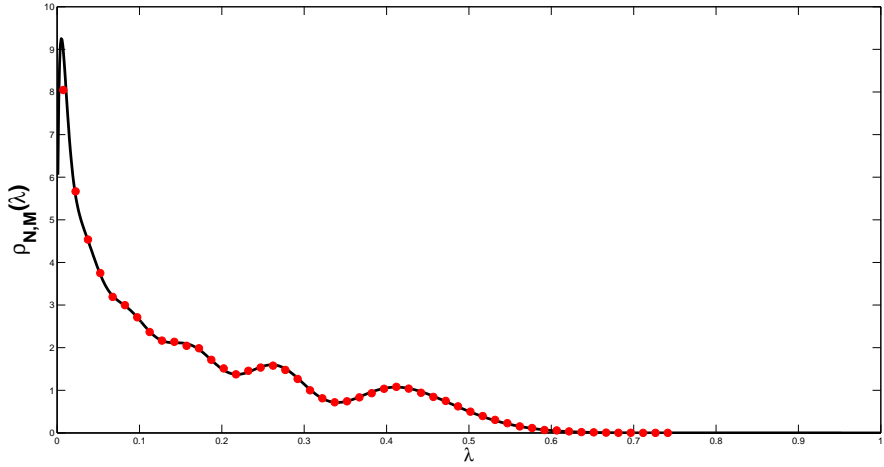}
\caption{Density of Schmidt eigenvalues for $\beta=1$ (eq. (\ref{rhofinale}), solid line) for $N=6,M=8$ compared with numerical simulations (red dots).\label{Beta1num}}
\end{center}
\end{figure}

In fig. \ref{Bothbetas} we plot the density (\ref{rhofinale}) for $\beta=1,2$ for $N=M=6$, and in fig. \ref{Beta1} the density (\ref{rhofinale})
for $N=6,M=12$ together with the large $N$ density (\ref{averagec}). In fig. \ref{Beta1num} we compare the theoretical density with numerical results, obtained as follows \cite{ZyckBook,ZS}:
\begin{enumerate}
\item we generate $n=5\cdot 10^4$ {\em real} Gaussian $M\times N$ matrices $\mathcal{X}$ (where $N=6,M=8$).
\item for each instance we construct the
Wishart matrix $\mathcal{W}=\mathcal{X}^T \mathcal{X}$. 
\item we diagonalize $\mathcal{W}$ and collect its $N$ real and non-negative eigenvalues $\{ \tilde{\lambda}_1,\ldots,\tilde{\lambda}_N\}$.
\item we define a new set of variables $0\leq \lambda_i\leq 1$ as $\lambda_i =\tilde{\lambda}_i /\sum_{i=1}^N \tilde{\lambda}_i$, for $i=1,\ldots,N$.
The set of variables $\lambda_i$ is guaranteed to be sampled according to the measure (\ref{jpddelta}).
\item we construct a normalized histogram of $\lambda_i$.
\end{enumerate}
The agreement between theory and simulations is excellent.

\section{Average R\'enyi entropy}\label{aver}
\begin{figure}[htb]
\begin{center}
\includegraphics[bb =0 0 240 154, width=0.7\textwidth]{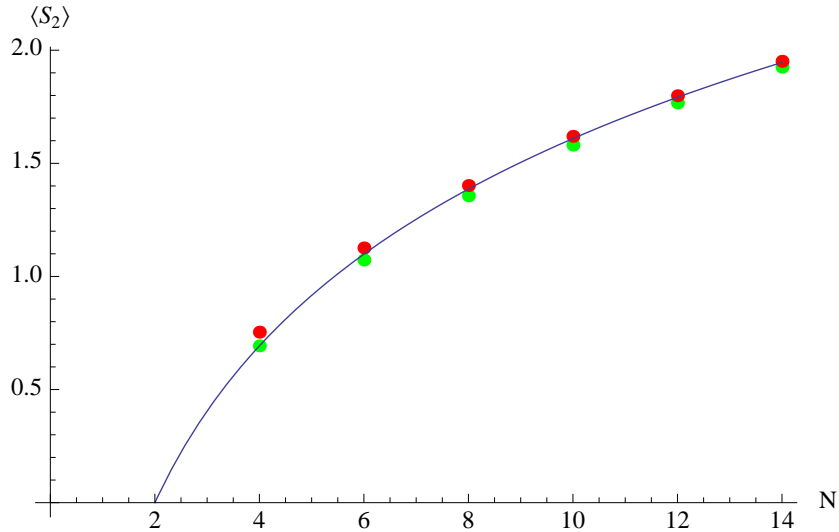}
\caption{Average purity $\langle\mathcal{S}_2\rangle$ as a function of $N(=M)$ for $\beta=1$ (green dots) and $\beta=2$ (red dots), compared with the exact asymptotic result eq. (\ref{gg})
for $N\to\infty$ \cite{majnadal}.\label{entropy1}}
\end{center}
\end{figure}

\begin{figure}[htb]
\begin{center}
\includegraphics[bb =0 0 240 156, width=0.7\textwidth]{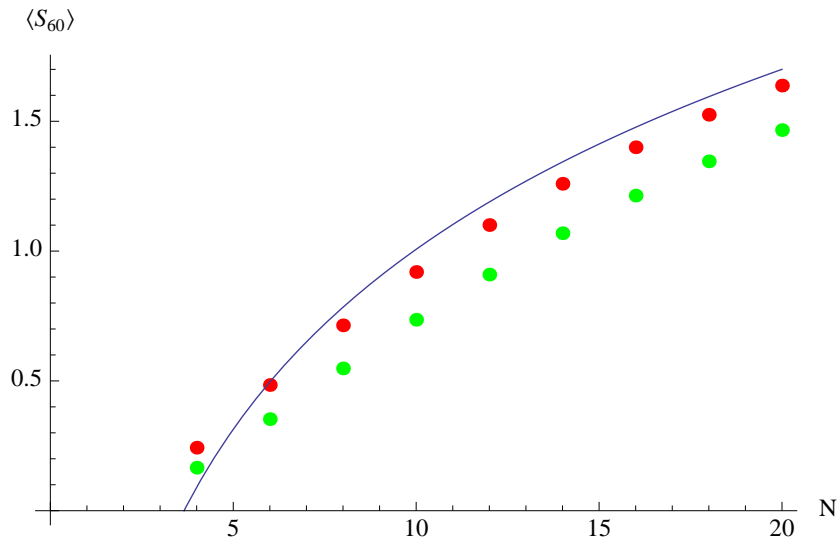}
\caption{Average R\'enyi entropy $\langle\mathcal{S}_{60}\rangle$ ($q=60$) as a function of $N(=M)$ for $\beta=1$ (green dots) and $\beta=2$ (red dots), compared with the exact asymptotic result eq. (\ref{gg})
for $N\to\infty$ \cite{majnadal}.\label{entropy2}}
\end{center}
\end{figure}

The average R\' enyi entropy $\langle\mathcal{S}_q\rangle$ is computed from the density (\ref{rhofinale}) as:
\begin{equation}\label{exactbeta1}
\langle\mathcal{S}_q\rangle=\frac{1}{1-q}\log\left[N\underbrace{\int_0^1 dx\ x^q\varrho_{N,M}(x)}_{\mathcal{J}(q)}\right]
\end{equation}
The integral $\mathcal{J}(q)$ can be computed easily. First, define the function:
\begin{equation}
\varphi(a,b,c,d):=\int_0^1 dz\ z^a (1+z)^b \mathrm{B}(z,c,d)
\end{equation}
Then the sought formula for $\mathcal{J}(q)$ reads:
\begin{equation}
\mathcal{J}(q) =\mathcal{N}_{N,M}\sum_{j=0}^{N-2}\sum_{m=0}^j\sum_{\ell=0}^{j+1} \mathbf{c}_{\ell m}^{(j)}
\left[\mathbf{\theta}_\ell(m+\ell,q)-\mathbf{\theta}_m(m+\ell,q)\right]
\end{equation}
where:
\begin{eqnarray}
\nonumber &\mathbf{\theta}_r(\alpha,q) :=\mathrm{B}\left(\frac{\nu-1}{2}+r+1,-\kappa-\alpha\right)\mathrm{G}_r(\alpha,q)+\\
& \nonumber + 2\varphi\left(q+\frac{\nu+1}{2}+\alpha-1-r,\frac{\nu-1}{2}+r-\kappa-\alpha,\frac{\nu-1}{2}+r+1,-\kappa-\alpha\right).\\		
&\nonumber\mathrm{G}_r(\alpha,q) :=
\mathrm{B}\left(q+\alpha -r+\frac{\nu+1}{2},\frac{\nu-1}{2}+r-\kappa-\alpha+1\right)-\\
&\nonumber\frac{2}{q+\alpha -r+\frac{\nu+1}{2}}\ _2 F_1 \left(q+\alpha -r+\frac{\nu+1}{2},
q+\nu-\kappa+1;q+\alpha -r+\frac{\nu+1}{2}+1;-1\right)
\end{eqnarray}
Here, $_2 F_1(a,b;c;x)$ is a hypergeometric function defined by the series:
\begin{equation}
_2 F_1(a,b;c;x)=\sum_{k=0}^\infty \frac{(a)_k (b)_k}{(c)_k\ k!}x^k
\end{equation}
In fig. \ref{entropy1} and \ref{entropy2} we compare respectively $\langle\mathcal{S}_2\rangle$ (average purity) and $\langle\mathcal{S}_{60}\rangle$ 
as a function of $N(=M)$ for $\beta=1,2$ with the large $N$ asymptotic formula (\ref{gg}) from \cite{majnadal}. 

We find that:
\begin{enumerate}
\item The average R\'enyi entropy for systems with time-reversal symmetry $(\beta=1)$ is always lower than systems of the same size where this symmetry is broken $(\beta=2)$.
This fact is in agreement with recent findings \cite{ArulSub} about so-called 'single-particle' or one-magnon states, where real states have lower entanglement measured in terms
of two-spin entanglement content than the case of complex states. We have checked that this feature persists for $N\neq M$, where a large $N$ formula for general $q$ is not yet available. 
\item For low $q$, the finite and large $N$ results are in excellent agreement already for $N\sim 6$, for both $\beta=1,2$. 
This means that for the most relevant cases of average von Neumann entropy $(q\to 1)$
and purity $(q=2)$, one can safely use eq. (\ref{gg}) from \cite{majnadal} as an excellent approximation for any practical purposes. Conversely, the quality of the approximation decays as $q$ increases,
up to the limit $\mathcal{S}_{\infty}\to -\ln\lambda_{\mathrm{max}}$ (where $\lambda_{\mathrm{max}}$ is the largest Schmidt eigenvalue), and one has to consider larger and larger  
subsystems in order to reach a satisfactory agreement (see fig. \ref{entropy2}). The discrepancy between the $\beta=1$ and $\beta=2$ is also more pronounced in the case of high $q$, and the
convergence to the asymptotic limit is much slower for the $\beta=1$ case.
\end{enumerate}

\section{Conclusions}\label{concl}
In summary, we have computed exactly the density $\varrho_{N,M}(\lambda)$ of Schmidt eigenvalues for bipartite entanglement of random pure states with orthogonal (time-reversal) symmetry $(\beta=1)$. The 
result is valid for any finite dimensions $N\leq M$ (with $N$ even) of the corresponding Hilbert space partitions. Using the exact formula we derived and a simple linear statistics, we compute
the average R\'enyi entropy $\langle\mathcal{S}_q\rangle$ for the $\beta=1$ case, which was previously unavailable. We find that the exact values for the averages at $N=M$ converge
very quickly to the asymptotic $N\to\infty$ formula derived in \cite{majnadal} for low values of the parameter $q$, thus including the most relevant cases of the von Neumann entropy $(q\to 1)$ and
the so-called \emph{purity} $(q=2)$. As $q$ is increased, the speed of convergence deteriorates for both $\beta=2$ and $\beta=1$, and the latter value for the average R\'enyi entropy is
consistently lower than the former for the same values of parameters $q,N,M$, even at $N\neq M$.\\
\vspace{10pt}

{\bf Acknowledgments.}
I am indebted and grateful to Gernot Akemann and Michael Phillips for helping me out with the Wishart-Laguerre density formula and for useful correspondence. I warmly thank
C\'eline Nadal, Satya Majumdar and Antonello Scardicchio for collaborations on related project and many interesting discussions, and Valerio Cappellini and Fabio Caccioli 
for a careful reading of the manuscript
and helpful advice.

\end{document}